\magnification=\magstep1
\hfuzz=6pt
\baselineskip=15pt

$ $

\vskip 1in
\centerline{\bf Quantum computation over continuous variables}

\bigskip
\centerline{Seth Lloyd}

\centerline{MIT Department of Mechanical Engineering}

\centerline{MIT 3-160, Cambridge, Mass. 02139, USA}

\bigskip
\centerline{\it and}

\bigskip
\centerline{Samuel L.\ Braunstein}

\centerline{SEECS, University of Wales}

\centerline{Bangor LL57 1UT, UK}

\bigskip
\noindent {\it Abstract:}  This paper provides
necessary and sufficient conditions for constructing
a universal quantum computer over continuous variables.  As an example, it
is shown how a universal quantum computer for the amplitudes of 
the electromagnetic field might be constructed using simple linear 
devices such as beam-splitters and phase shifters, together 
with squeezers and nonlinear devices such as Kerr-effect fibers 
and atoms in optical cavities.  Such a device could in principle perform
`quantum floating point' computations.  Problems of noise, finite 
precision, and error correction are discussed.

\bigskip
Quantum computation has traditionally concerned itself
with the manipulation of discrete systems such as 
quantum bits, or `qubits'.$^{1-2}$  Many quantum variables,
such as position and momentum, or the amplitudes of electromagnetic
fields, are continuous.  Although noise and finite precision
make precise manipulations of continuous variables intrinsically
more difficult than the manipulation of discrete variables,
because of the recent developments in quantum error correction$^{3-5}$
and quantum teleportation$^{6-7}$ of continuous quantum variables 
it is worthwhile addressing the question of 
quantum computation over continuous variables.  

At first it might seem that quantum computation over continuous
variables is an ill-defined concept.  First consider quantum computation
over discrete variables.
A universal quantum computer over discrete variables such as 
qubits can be defined to be a device that can by local operations
perform any desired unitary transformation over those variables.$^{1-2,8}$  
More precisely, a universal quantum computer applies `local' operations 
that effect only a few variables at a time (such operations are called
quantum logic gates): by repeated application of such local operations
it can effect any unitary transformation over a finite number of 
those variables to any desired degree of precision. 
Now consider the continuous case.  Since an arbitrary unitary transformation
over even a single continuous variable requires an infinite
number of parameters to define, it typically cannot be approximated
by any finite number of continuous quantum operations such
as, for example, the application of beam-splitters, phase-shifters,
squeezers, and nonlinear devices to modes of the electromagnetic
field.  It is possible, however, to define a notion of universal 
quantum computation over continuous variables for various subclasses
of transformations, such as those that correspond to Hamiltonians
that are polynomial functions of the operators corresponding to
the continuous variables: a set of continuous quantum operations
will be termed universal for a particular set of transformations
if one can by a finite number of applications
of the operations approach arbitrarily closely to any transformation
in the set.

This paper provides necessary and sufficient conditions for 
universal quantum computation over continuous variables for
transformations that are polynomial in those variables.
Such a continuous quantum computer is shown to be capable
in principle of performing arithmetical manipulations of continuous
variables in a `quantum floating point' computation.
Issues of noise and finite precision are discussed and
applications proposed.

Consider a single continuous variable corresponding to an
operator $X$.  Let $P$ be the conjugate variable:
$[X,P]=i$.  For example, $X$ and $P$ could correspond to 
quadrature amplitudes of a mode of the electromagnetic field 
(the quadrature amplitudes are the real and imaginary parts of
the complex electric field).
First investigate the problem of constructing Hamiltonians
that correspond to arbitrary polynomials of $X$ and $P$.
It is clearly necessary that one be able to 
apply the Hamiltonians $\pm X$ and $\pm P$ themselves.
In the Heisenberg picture, applying a Hamiltonian $H$ gives
a time evolution for operators $\dot A = i[H,A]$, so that
$A(t)=e^{iHt}A(0)e^{-iHt}$.
Accordingly, applying the Hamiltonian $X$ for time $t$ 
takes $X \rightarrow X, P \rightarrow P - t$, and applying
$P$ for time $t$ takes $X\rightarrow X + t, P\rightarrow P$:
the Hamiltonians $X$ and $P$ have the effect of shifting the 
conjugate variable by a constant.  In the case of the
electromagnetic field, these Hamiltonians correspond to 
linear displacements or translations of the quadrature amplitudes. 

To construct an arbitrary Hamiltonian of the form $aX
+bP + c$, first apply $X$ for a time $a dt$, where $dt$ is
a short period of time, then apply $P$ for a time $bdt$,
and finally apply $P$ for a time $\sqrt{ c dt}$, 
 $X$ for a time $\sqrt{ c dt}$,
 $-P$ for a time ${\sqrt c dt}$,
 and $-X$ for a time $\sqrt{ c dt}$.  The net effect is
a transformation 
$$\eqalign{e^{-iX\sqrt{cdt}}e^{-iP\sqrt{cdt}}&e^{iX\sqrt{cdt}}
e^{iP\sqrt{cdt}} e^{ibPdt} e^{iaXdt} \cr
&= 1 + i\big( -i[X, P] c + bP + aX\big)dt + {\rm O}(dt^{3/2})\cr
&\approx e^{i(aX+bP+c)dt}\cr}
\eqno(1)$$
\noindent By making $dt$ sufficiently small, one can approach
arbitrarily closely to effecting a Hamiltonian of the desired form
over small times.  By repeating the small-time construction $t/dt$
times, one can approach arbitrarily closely to effecting the
desired Hamiltonian over time $t$.

There are clearly simpler ways to enact an overall phase shift.
When applied to arbitrary sets of Hamiltonians, however,
the construction given above provides the prescription for determining
exactly what Hamiltonians can be constructed by the repeated
application of Hamiltonians from the set:
if one can apply a set of Hamiltonians $\{\pm H_i\}$, one can 
construct any Hamiltonian that is a linear combination of
Hamiltonians of the form
$\pm i[H_i,H_j]$, $\pm [H_i,[H_j,H_k]]$, etc.,$^{9-13}$ 
and no other Hamiltonians.  That is, one can
construct the Hamiltonians in the algebra generated from the original 
set by commutation.  This key point, originally derived in the
context of quantum control and discrete quantum
logic, makes it relatively simple to determine the set of Hamiltonians
that can be constructed from simpler operations.

The application of the translations $\pm X$ and $\pm P$ 
for short periods of time
clearly allows the construction of any Hamiltonian that 
is linear in $X$ and $P$; this is all that it allows.  Suppose
now that one can apply the quadratic Hamiltonian $H = 
(X^2+P^2)/2$.  Since $\dot P = i[H,P]=X$, $\dot X = i[H,X]= - P$,
application of this Hamiltonian for time $t$ takes
$X\rightarrow \cos t X - \sin t P$, $P\rightarrow \cos t P
+ \sin t X$.  If $X$ and $P$ are quadrature amplitudes of
a mode of the electromagnetic field, then $H$ is just the Hamiltonian
of the mode (with frequency $\omega=1$) and corresponds to a phase shifter.
Hamiltonians of this form can be enacted by letting the system
evolve on its own or by inserting artificial phase delays. 
Note that since $e^{iHt}$ is periodic with period $1/4\pi$, 
one can effectively apply $-H$ for a time $\delta t$
by applying $H$ for a time $4\pi -\delta t$.  The simple
commutation relations between $H,X$ and $P$ imply that the addition
of $\pm H$ to the set of operations that can be applied
allows the construction of Hamiltonians of the form
$aH+bX+cP+d$.

Suppose that in addition to translations and phase shifts one can
apply the quadratic Hamiltonian $\pm S = \pm(XP+PX)/2$.  
$S$ has the effect $\dot X=i[S,X]=X$, 
$\dot P=i[S,P]=-P$, i.e., applying $+S$ takes $X\rightarrow e^t X$,
$P\rightarrow e^{-t}P$: $S$ `stretches' $X$ and `squeezes' $P$ by
some amount. Similarly $-S$ squeezes $X$ and stretches $P$.
In the terminology of quantum optics, $S$ corresponds
to a squeezer operating in the linear regime.  It can easily be verified that 
$[H,S]=i(X^2-P^2)$.  Looking at the algebra generated from
$X,P,H$ and $S$ by commutation, one sees that translations,
phase shifts, and squeezers allow the construction of any 
Hamiltonian that is quadratic in $X$ and $P$, and of no Hamiltonian
of higher order.  

To construct higher order Hamiltonians, nonlinear operations are
required.  One such operation is the `Kerr' Hamiltonian
$H^2=(X^2+P^2)^2$, corresponding to a $\chi^3$ process in nonlinear
optics.  This higher order Hamiltonian has the key
feature that whereas commuting the previous Hamiltonians, $X,P,H,S$
with some polynomial in $X$ and $P$ resulted in a polynomial with
the same or lower order, commuting $H^2$ with a polynomial in
$X$ and $P$ typically {\it increases} its order, e.g.,
$$\eqalignno{&[H^2, X] = H[H,X]+[X,H]H= i(HP+PH)=
i(X^2P+ PX^2 + 2P^3)/2&(2.1)\cr
&[H^2,P] = -i(P^2X+XP^2+2X^3)/2&(2.2)\cr
&[H^2,S]=H[H,S]+[H,S]H=i(X^4-P^4)/2 \quad.&(2.2)\cr}$$
\noindent By evaluating a few more commutators, e.g.,
$[X,[H^2,S]]=P^3$, $[P,[H^2,S]]=X^3$ one sees that the algebra generated by
$X,P,H,S$ and $H^2$ by commutation includes 
all third order polynomials in $X$ and $P$.  A simple inductive
proof now shows that one can construct Hamiltonians that are arbitrary
Hermitian polynomials in any order of $X$ and $P$.   Suppose that
one can construct any polynomial of order $M$ or less, where
$M$ is of degree at least 3.
Then since $[P^3, P^mX^n]=iP^{m+2}X^{n-1} +$ lower order terms,
and  $[X^3, P^mX^n]=iP^{m-1}X^{n+2} +$ lower order terms, 
one can by judicious commutation of $X^3$ and $P^3$ with monomials of 
order $M$ construct any monomial of order $M+1$.  Since any polynomial
of order $M+1$ can be constructed from monomials of order $M+1$ and
lower, by applying linear operations and a single
nonlinear operation a finite number of times one can
construct polynomials of arbitrary order in $X$ and $P$
to any desired degree of accuracy.   Comparison with similar
results for the discrete case$^{14}$ shows that the number 
of operations required grows as a small polynomial in the order of 
the polynomial to be created, the accuracy to which that polynomial
is to be enacted, and the time over which it is to be applied. 

The use of the Kerr Hamiltonian $H^2$ was not essential:
any higher order Hamiltonian will do the trick.
Note that commutation of a polynomial in $X$ and $P$ with
$X$ and $P$ themselves (which have order 1) always reduces the order 
of the polynomial by at least 1, commutation with $H$ and 
$S$ (which have order 2) never increases the order, and 
commutation with a polynomial of order 3 or higher typically 
increases the order by at least 1.  Judicious
commutation of $X,P,H$ and $S$ with an applied Hamiltonian
of order 3 or higher therefore allows the construction of
arbitrary Hermitian polynomials of any order in $X$ and $P$.

The above set of results shows that simple linear operations,
together with a single nonlinear operation, allow one to construct
arbitrary polynomial Hamiltonian transformations of a single quantum variable.
Let us now turn to more than one variable, e.g., the case of an
interferometer in which many modes of the electromagnetic
field interact.  Suppose now that there are many variables, 
$\{X_i,P_i\}$, on each of which the simple single-variable operations 
described above can be performed.  Suppose in addition
Hamiltonians of the form $\pm B_{ij}= \pm (P_iX_j-X_iP_j)$
can be applied.  Since $\dot X_i= i[B_{ij},X_i]= X_j$
 $\dot X_j= i[B_{ij},X_j]= -X_i$,
 $\dot P_i= i[B_{ij},P_i]= P_j$,
 $\dot P_j= i[B_{ij},P_j]= -P_i$,
this operation has the effect of taking 
$A_i\rightarrow \cos t A_i + \sin t A_j$,
$A_j\rightarrow \cos t A_j - \sin t A_i$, for $A_i=X_i,P_i$, 
$A_j=X_j,P_j$.  For the
electromagnetic field, $B_{ij}$ functions as a beam
splitter, linearly mixing together the two modes $i$ and $j$.
By repeatedly taking commutators of $B_{ij}$ with polynomials
in $X_i,P_i$, it can be easily seen by the same algebraic arguments
as above that it is possible to build up arbitrary Hermitian polynomials
in $\{X_i, P_i \}$.

This concludes the derivation of the main result: simple linear
operations such as translations, phase shifts, squeezers, and
beam splitters, combined with some nonlinear operation such 
as a Kerr nonlinearity, suffice to enact to an arbitrary degree
of accuracy Hamiltonian operators that are arbitrary polynomials
over a set of continuous variables.  Note that in contrast to
the case of qubits, in which a nonlinear coupling between qubits
is required to perform universal quantum computation, in the
continuous case only {\it single variable} nonlinearities
are required, along with linear couplings between the variables.  

In analog with information over classical continuous variables,
which is measured in units of `nats' (1 nat = ${\rm log}_2 e$
bits), the unit of continuous quantum information will be called
the `qunat.'  
Two continuous variables in the pure state $|\psi\rangle_{12}$ 
possess $-{\rm tr} \rho_1 \ln \rho_1$ qunats of entanglement,
where $\rho_1 = {\rm tr}_2 |\psi\rangle_{12}\langle\psi|$.
For two squeezed vacua (squeezed by an amount $e^{-r}$) entangled
using a beam splitter as in refs. (5-7) the
entropy so computed from the approximate EPR state is given by
$$
S(\rho)=(1+\bar n)\ln (1+\bar n) - \bar n \ln \bar n \, ~~ {\rm qunats}
\eqno(3)
$$
\noindent with $\bar n = e^r \sinh r$. 
For example, $e^{2r} = 10$ gives 10 dB of squeezing in power, corresponding
to $r = 1.15129$.  By equation 3, two continuous variables entangled
using a 10 dB squeezer then possess 2.60777 qunats of shared, continuous
quantum information, equivalent to 3.76221 qubits of discrete quantum
information.

Quantum computation over continuous variables can be thought
of as the systematic creation and manipulation of qunats.
Universal quantum computation for
polynomial transformations of continuous variables effectively allows
one to perform `quantum floating point' manipulations on those
variables.  For example, it is clearly possible using
linear operations alone to take the inputs $X_1, X_2$ and to map
them to $X_1, aX_1+bX_2 +c$.  Similarly, application of the
three-variable Hamiltonian $X_1X_2P_3$ allows one to multiply
$X_1$ and $X_2$ and place the result in the `register' $X_3$:
$$\eqalign{
&\dot X_1=  i[X_1X_2P_3, X_1] = 0, 
\dot X_2= i[X_1X_2P_3, X_3] =0, \dot X_3 = i[X_1X_2P_3, X_3]= X_1X_2\cr
&X_1\rightarrow X_1, X_2\rightarrow X_2,
X_3\rightarrow X_3 + X_1X_2t.\cr}\eqno(4)$$  
A wide variety of quantum floating point operations are
possible.  Any polynomial transformation of the continuous
variables is clearly possible, as is any transformation that
can be infinitesimally represented by a convergent power series.
Just as classical computation 
over continuous variables in principle allows one to solve 
problems more rapidly than is possible digitally,$^{14}$ 
it is interesting to speculate that quantum computation
over continuous variables might in principle allow the solution of
problems more rapidly than is possible using a `conventional,' discrete
quantum computer.  Continuous variable computation has its own set
of problems that might be sped up by the application of continuous
quantum computation: for example, such a continuous quantum
computer might be used to investigate continuous $NP$-complete problems
such as the 4-Feasibility problem, that is, the problem of deciding whether
or not a real degree 4 polynomial in $n$ variables has a zero.$^{15}$
In practice, of course, due to finite precision
a continuous quantum computer will effectively be able to solve
the same set of problems that a `conventional' discrete quantum
computer can, although it may be able to perform some operations
more efficiently.

The ability to create and manipulate qunats depends
crucially on the strength of squeezing and of the nonlinearities
that one can apply.  10 dB squeezers (6 dB after attenuation in
the measurement apparatus) currently exist.$^{16}$ 
High Q cavity quantum electrodynamics can supply a strong Kerr effect in 
a relatively lossless context, and quantum logic gates constructed for qubits
could be used to provide the nonlinearity for continuous quantum
variables as well.$^{17}$  Here the fact that only single-mode nonlinearities
are required for universal quantum computation simplifies the
problem of effecting continuous quantum logic.  Nonetheless,
the difficulty of performing repeated nonlinear operations
in a coherent and loss-free manner is likely to limit the possibilities
for quantum computation over the amplitudes of the electromagnetic
field. 

Noise poses a difficult problem for quantum computation,$^{18-20}$  
and continuous variables
are more susceptible to noise than discrete variables.
Since an uncountably infinite number of things can go wrong with a continuous
variable, it might at first seem that continuous error correction
routines would require infinite redundancy.  In fact, continuous
quantum error correction routines exist and require no greater
redundancy than conventional routines.$^{3-5}$  
Such routines are capable
of correcting for noise and decoherence in principle: in practice,
measurement noise, losses, and the lack of perfect squeezing will lead to
imperfect error correction.$^5$  Surprisingly, continuous
quantum error correction routines are in some sense easier
to enact than discrete quantum error correction routines,
in that the continuous routines can be implemented using
only {\it linear} operations together with classical feedback.$^5$
The relative simplicity of such routines suggests that robust,
fault-tolerant quantum computation may in principle be possible for continuous
quantum variables as well as for qubits (A scheme for quantum computation
is fault-tolerant if quantum computations can be carried out 
even in the presence of noise and errors.$^{21-22}$ A fault-tolerant
scheme that allows for arbitrarily long quantum computations to
be carried out is said to be robust.$^{23}$).
If this is indeed the case then quantum computation over continuous variables,
despite its intrinsic difficulties, may be an experimentally
viable form of quantum information processing.  Continuous variables 
might be used to simulate continuous quantum systems such as quantum field
theories.  Even in the absence of fault tolerance,
the large bandwidths available to continuous
quantum computation make it potentially useful for 
quantum communications and cryptography.$^{24}$  
\vfill
\noindent{\it Acknowledgements:} S.L. would like to thank H.\ Haus
and H.\ J.\ Kimble for useful discussions.
\vfil\eject

\centerline{\bf References}

\bigskip
\noindent 1. D.\ DiVincenzo, {\it Science\/} {\bf 270}, 255 (1995).

\noindent 2. S.\ Lloyd, {\it Sci.\ Am.\/} {\bf 273}, 140 (1995).

\noindent 3. S.\ Lloyd and J.\ J.-E.\ Slotine, {\it Phys.\ Rev.\ Lett.\/}
{\bf 80}, 4088 (1998).

\noindent 4. S.\ L.\ Braunstein, {\it Phys.\ Rev.\ Lett.\/}
{\bf 80}, 4084 (1998).

\noindent 5. S.\ L.\ Braunstein, {\it Nature\/} {\bf 394}, 47
(1998).

\noindent 6. S.\ L.\ Braunstein and H.\ J.\ Kimble, {\it Phys.\ Rev.\ Lett.\/}
{\bf 80}, 869 (1998).

\noindent 7. A.\ Furusawa, {\it et al}, {\it Science\/} {\bf 282}
706 (1998).

\noindent 8. This definition of quantum computation corresponds to
the normal `circuit' definition of quantum computation as in, e.g.,  
D.\ Deutsch, {\it Proc.\ Roy.\ Soc.\ A}, {\bf 425}, 73 (1989), and
A.\ C.-C.\ Yao, in
{\it Proceedings of the 36th Annual Symposium on Foundations
of Computer Science}, S.\ Goldwasser, Ed., IEEE Computer
Society, Los Alamitos, CA, 1995, pp. 352-361.
The work of M.\ Reck {\it et al.},
{\it Phys.\ Rev.\ Lett.\/} {\bf 73}, 58 (1994), and of    
N.J. Cerf, C. Adami, and P.G. Kwiat, {\it Phys. Rev. A}, {\bf 57}
R1477 (1998), showing how to
perform arbitrary unitary operators using only linear devices
such as beam splitters, though of considerable interest and potential
practical importance, does not constitute quantum computation
by the usual definition.  Reck {\it et al.} and 
Cerf {\it et al.} propose performing arbitrary unitary operations
on $N$ variables not by acting on the variables themselves but
by expanding the information in the variables into an interferometer
with $O(2^N)$ arms and acting in this exponentially larger space.
Local operations on the original variables correspond to highly
nonlocal operations in this `unary' representation: to flip a single
bit requires one to act on half ($O(2^{N-1})$) of the arms of the
interferometer.  Actually to perform quantum computation on qubits using
an interferometer requires nonlinear operations as detailed in
Y.\ Yamamoto, M.\ Kitagawa, and K.\ Igeta, 
in {\it Proceedings of the 3rd Asia-Pacific
Physics Conference}, Y.\ W.\ Chan, A.\ F.\ Leung, C.\ N.\ Yang, K.\ Young,
eds., World Scientific, Singapore, 1988, pp. 779-799, and
G.\ J.\ Milburn, {\it Phys.\ Rev.\ Lett.\/} {\bf 62} 2124 (1989).

\noindent 9. G.\ M.\ Huang, T.\ J.\ Tarn, J.\ W.\ Clark, On the
controllability of quantum-mechanical systems.
{\it J.\ Math.\ Phys.\/} {\bf 24}(11), 2608-2618 (1983).

\noindent 10. R.\ W.\ Brockett, R.\ S.\ Millman, H.\ J.\ Sussman,
eds., {\it Differential Geometric Control Theory\/}
(Birkhauser, Boston, 1983).  Z.\ Li, J.\ F.\ Canney, eds., {\it
Nonholonomic Motion Planning\/} (Kluwer Academic, Boston,
1993).

\noindent 11. V.\ Ramakrishna, M.\ V.\ Salapaka, M.\ Dahleh, H.\ Rabitz,
A.\ Peirce, {\it Phys.\ Rev.\ A\/} {\bf 51}, 960-966 (1995).

\noindent 12. S.\ Lloyd, {\it Phys.\ Rev.\ Lett.\/}, {\bf 75},
346-349 (1995).

\noindent 13. D.\ Deutsch, A.\ Barenco, A.\ Ekert, {\it Proc.\
Roy.\ Soc.\ A}, {\bf 449}, 669-677 (1995).

\noindent 14. S.\ Lloyd, {\it Science\/} {\bf 273}, 1073 (1996).

\noindent 15. L.\ Bluhm, M.\ Shub, and S.\ Smale, {\it Bull.\ Am.\ Math.\
Soc.\/} {\bf 21}, 1-46 (1989).

\noindent 16. L.\ A.\ Wu {\it et al.}, {\it Phys.\ Rev.\ Lett.\/} {\bf 57},
2520 (1986).

\noindent 17. Q.\ A.\ Turchette, {\it et al},
{\it Phys.\ Rev.\ Lett.\/}, {\bf 75}, pp. 4710-4713 (1995).

\noindent 18. R.\ Landauer, {\it Nature\/}
{\bf 335}, 779-784 (1988).

\noindent 19. R.\ Landauer,
{\it Phys.\ Lett.\ A}, {\bf 217}, 188-193 (1996).

\noindent 20. R.\ Landauer,
{\it Phil. Trans. Roy. Soc.
Lond. A\/} {\bf 335}, 367-376 (1995).

\noindent 21. P.\ Shor, 
{\it Proceedings of the 37th Annual Symposium on the Foundations
of Computer Science,} IEEE Computer Society Press, Los Alamitos,
1996, pp. 56-65.  

\noindent 22. D.\ P.\ DiVincenzo and P.\ W.\ Shor, {\it
Phys.\  Rev.\ Lett.\/} {\bf 77}, 3260-3263 (1996).

\noindent 23. R.\ Laflamme, M.\ Knill, W.H. Zurek, {\it
Science}  {\bf 279}, 342 (1998).   D.\ Aharanov and Ben-Or, 
quant-ph.  J.\ Preskill, {\it Proc. Roy. Soc. Lond. Ser. A} {\bf 454},
385 (1998).

\noindent 24. C.\ H.\ Bennett and G.\ Brassard, in {\it Proceedings
of the IEEE International Conference on Computers, Systems,
and Signal Processing, Bangalore, India}, IEEE Press, New York,
1984, pp. 175-179.
A.\ K.\ Ekert, {\it et al}, {\it Phys.\ Rev.\ Lett.\/} {\bf 69} 1293 (1992).
P.\ D.\ Townsend, J.\ G.\ Rarity and P.\ R.\ Tapster,
{\it Electronics Letters\/} {\bf 29}, 1291 (1993).
R.\ J.\ Hughes, {\it et al}, in {\it Advances in Cryptology: Proceedings of
Crypto 96}, Springer-Verlag, New York, pp. 329-343 (1997).
A.\ Muller {\it et al.} {\it Appl. Phys. Lett.}
{\bf 70}, 793 (1997).

\vfill\eject\end